\documentclass[12pt,preprint]{aastex}

\shorttitle{The Mass of SMM\,J02399-0136}
\shortauthors{Genzel et al.}

\received{2002 August 22}
\begin{document}

\title{Spatially Resolved Millimeter Interferometry of SMM\,J02399-0136: a 
Very Massive Galaxy at $z = 2.8$\altaffilmark{1}}

\altaffiltext{1}{Based on observations obtained at the IRAM Plateau de Bure 
interferometer.  IRAM is funded by the INSU/CNRS (France), the MPG (Germany), 
and the IGN (Spain).}

\author{Reinhard Genzel\altaffilmark{2,3}, Andrew J. Baker\altaffilmark{2}, 
Linda J. Tacconi\altaffilmark{2}, Dieter Lutz\altaffilmark{2}, Pierre 
Cox\altaffilmark{4}, St{\' e}phane Guilloteau\altaffilmark{5,6}, \& Alain 
Omont\altaffilmark{7}}

\altaffiltext{2}{Max-Planck-Institut f{\" u}r extraterrestrische Physik,
Postfach 1312, D-85741 Garching, Germany}

\altaffiltext{3}{Department of Physics, 366 Le~Conte Hall, University of 
California, Berkeley, CA 94720-7300}

\altaffiltext{4}{Institut d'Astrophysique Spatiale, Universit{\' e} de Paris 
XI, F-91405 Orsay, France}

\altaffiltext{5}{Institut de Radio Astronomie Millim{\' e}trique, 300 rue de 
la Piscine, F-38406 Saint Martin d'H{\` e}res, France}

\altaffiltext{6}{European Southern Observatory, Karl-Schwarzschild-Strasse 2, 
D-85741 Garching, Germany}

\altaffiltext{7}{Institut d'Astrophysique de Paris, CNRS, 98bis boulevard 
Arago, F-75014 Paris, France}

\begin{abstract}
We report high-resolution millimeter mapping with the IRAM Plateau de Bure 
interferometer of rest-frame $335\,{\rm \mu m}$ continuum and CO(3--2) line 
emission from the $z = 2.8$ submillimeter galaxy SMM\,J02399-0136.  The 
continuum emission comes from a $\sim 3\arcsec$ diameter structure whose 
elongation is approximately east-west and whose centroid is coincident within 
the astrometric errors with the brightest X-ray and rest-UV peak (L1).  The 
line data show that this structure is most likely a rapidly rotating disk.  
Its rotation velocity of $\geq 420\,{\rm km\,s^{-1}}$ implies a total 
dynamical mass of $\geq 3 \times 10^{11}\,{\rm sin}^{-2}\,i\,h_{0.7}^{-1}\,
M_\odot$ within an intrinsic radius of $8\,h_{0.7}^{-1}\,{\rm kpc}$, most of 
which is plausibly in the form of stars and gas.  SMM\,J02399-0136 is thus a 
very massive system, whose formation at $z \sim 3$ is not easy to understand 
in current CDM hierarchical merger cosmogonies.
\end{abstract}


\section{Introduction}\label{s-intro}

The strength of the extragalactic far-IR/submillimeter background indicates 
that about half of the cosmic energy density comes from distant, dusty 
starbursts and AGN \citep{fixs98,haus98,laga99,pei99}.  Surveys with 
ISOCAM at $15\,{\rm \mu m}$, SCUBA at $850\,{\rm \mu m}$, and MAMBO at 
$1200\,{\rm \mu m}$ suggest that this background is dominated by luminous 
and ultraluminous infrared galaxies ([U]LIRGs: $L_{\rm IR} \sim 10^{11.5} - 
10^{13}\,L_\odot$) at $z \geq 1$ (e.g., Smail, Ivison, \& Blain 1997; Hughes 
et al. 1998; Barger, Cowie, \& Sanders 1999; Aussel et al. 1999, 2000; Carilli 
et al. 2001; Elbaz et al. 2002; Genzel \& Cesarsky 2000).  Faint submillimeter 
sources frequently have only weak (if any) counterparts in the rest-frame UV 
and optical, however \citep{smai00,smai02,dann02}.  Redshifts and 
spectroscopic parameters have thus far been confirmed with CO interferometry 
for only three of the $> 10^2$ detected systems (Blain et al. 2002, and 
references therein).  

One of these three sources is SMM\,J02399-0136 (hereafter J02399), the 
brightest of the 15 significant background sources discovered in the SCUBA 
cluster lens survey (SCLS: Smail et al. 1997, 2002).  This galaxy's $850\,{\rm 
\mu m}$ flux density of 23.0\,mJy includes a magnification factor of 2.45 due 
to the $z = 0.37$ foreground cluster Abell\,370 (Ivison et al. 1998; R. 
Ivison, private communication).  For a dust temperature of 45\,K and 
emissivity index $\beta = 1.5$ (typical for local luminous {\it IRAS} 
galaxies: Dunne et al. 2000), its intrinsic IR luminosity (from $8-1000\,{\rm 
\mu m}$ in the rest frame) is $L_{\rm IR} = 1.2 \times 10^{13}\,h_{0.7}^{-2}\,
L_\odot$.\footnote{This paper assumes a flat $\Omega_\Lambda = 0.7$ cosmology 
with $H_0 = 70\,h_{0.7}\,{\rm km\,s^{-1}\,Mpc^{-1}}$, for which $z_{\rm CO} = 
2.808$ corresponds to a luminosity distance $D_{\rm L} = 23.5\,h_{0.7}^{-1}\,
{\rm Gpc}$ and an angular diameter distance $D_{\rm A} = 1.6\,h_{0.7}^{-1}\,
{\rm Gpc}$.}  J02399 is rich in gas as well as dust: \citet{fray98} detect 
strong CO(3--2) emission at $z_{\rm CO} = 2.808$, slightly redward of its UV 
line emission at $z_{\rm UV} = 2.803$ \citep{ivis98}.  The optical through 
X-ray properties of J02399 resemble those of local type 2 AGN and broad 
absorption line QSOs \citep{ivis98,vern01}.  Moderately broad ($2000\,{\rm 
km\,s^{-1}}$) ${\rm Ly}\,\alpha$ emission is extended over about $13\arcsec$, 
with a maximum on the brightest UV continuum peak L1, and a second maximum on 
the UV continuum peak L2 lying $\sim 3\arcsec$ farther east 
\citep{ivis98,vern01}.  \citet{baut00} find that strong X-ray emission from a 
compact nucleus is obscured by a column of $\sim 10^{24}\,{\rm cm^{-2}}$ 
and estimate an unabsorbed X-ray luminosity of $2.7 \pm 1.0 \times 
10^{44}\,h_{0.7}^{-2}\,{\rm erg\,s^{-1}}$.  The ratios of X-ray, far-IR, and 
radio fluxes suggest that 15--70\% of the bolometric (infrared) luminosity may 
come from the AGN \citep{baut00,fray98}, as is true for composite 
starburst/AGN members of the local ULIRG population like Mkn\,231 
\citep{sand96,genz98}.

\section{Observations}\label{s-obs}

We observed J02399 with the B, C, and D configurations of the IRAM Plateau de 
Bure interferometer (PdBI: Guilloteau et al. 1992) in November -- December 
1998 and in January -- September 2002.  In 1998, the array consisted of five 
15\,m telescopes, each equipped with both [single-sideband] 3\,mm and 
[double-sideband] 1\,mm SIS receivers which we used simultaneously.  System 
temperatures (referred to above the atmosphere) were $\sim 120\,{\rm K}$ and 
$\sim 400\,{\rm K}$ at 3\,mm and 1\,mm, respectively.  We tuned the 3\,mm 
receivers to 90.81\,GHz, the frequency of the redshifted CO(3--2) line for 
$z_{\rm CO} = 2.808$, and the 1\,mm receivers to a line-free region at 
235.00\,GHz corresponding to a rest wavelength of $335\,{\rm \mu m}$.  Four 
correlator modules were deployed to sample the full 536\,MHz of contiguous 
3\,mm bandwidth at $2.5\,{\rm MHz\,channel^{-1}}$ resolution, while the 
remaining two modules provided 295\,MHz of 1\,mm continuum bandwidth.  By 
2002, the array had expanded to six telescopes, eight correlator modules, and 
a wider IF bandwidth.  We therefore retuned the 1\,mm receivers to 
211.83\,GHz, the frequency of the redshifted CO(7--6) line, and deployed four 
correlator modules (giving 559\,MHz of continuous bandwidth) at both 3\,mm and 
1\,mm.  The pointing center was also changed by $0.91\arcsec$ for the 2002 
observations, to agree with the CO(3--2) centroid ($\alpha$ (J2000) $ = 02^{h} 
39^{m} 51.89^{s}$ and $\delta$ (J2000) $= -01^{\circ} 35^{\prime} 
59.9^{\prime\prime}$) measured by \citet{fray98}.  

We calibrated the 1998 and 2002 data separately using the CLIC routines in the 
IRAM GILDAS package \citep{guil00}.  Passband calibration used one or 
more bright quasars.  Phase and amplitude variations within each track were 
calibrated out by interleaving reference observations of one or two quasars 
near the source every 30 minutes.  The overall flux scale for each epoch was 
set by comparisons of model sources CRL618 and MWC349 with bright quasars 
whose flux densities are regularly monitored at both the PdBI and the IRAM 
30\,m telescope.  Once the source data were calibrated, we precessed the 1998 
visibilities to the 2002 phase center.  After confirming our nondetection of 
CO(7--6) emission, we rescaled the 2002 1\,mm continuum visibilities from 
211.83 to 235.00 GHz using a power law index of 3 derived from (sub)millimeter 
photometry \citep{ivis98,smai02}.  We then combined the 1998 and 2002 $uv$ 
data, yielding a total of 40 distinct baselines (24--331\,m) from the three 
array configurations, and smoothed the CO(3--2) data to $15\,{\rm 
km\,s^{-1}}$ resolution.  After flagging bad and high phase noise data, we 
were left with the [on-source, six-antenna array] equivalents of 39.6 hours of 
3\,mm data and 21.6 hours of 1\,mm data.  We deconvolved the data cubes with 
natural weighting using the CLEAN algorithm as implemented in the NRAO AIPS 
package \citep{vanm96} and GILDAS.  Although no 3\,mm (rest-frame $840\,{\rm 
\mu m}$) continuum emission was seen, both the CO(3--2) line and the 
[rest-frame] $335\,{\rm \mu m}$ continuum emission are well detected and 
resolved in our maps.  

\section{Results}\label{r-res}

\subsection{Continuum emission}

Figure \ref{f-cont} shows the $335\,{\rm \mu m}$ ($\lambda_{\rm obs} 
= 1.27\,{\rm mm}$) continuum emission from J02399 at a resolution of 
$1.8\arcsec \times 1.4\arcsec$, superposed on a sub-arcsecond resolution 
$R$-band image obtained at Keck (L. Cowie, private communication).  The 
observed total flux density of $7.0 \pm 1.2\,{\rm mJy}$ ($\pm 1.4\,{\rm 
mJy}$ for a 20\% uncertainty in the flux scale) agrees well with the value 
(8.5\,mJy) expected from JCMT photometry at 850 and $1350\,{\rm \mu m}$, 
suggesting that the interferometer has resolved out little if any emission.  
The central continuum structure is clearly resolved in the east-west 
direction, parallel to the dominant lensing shear predicted by the SCLS 
lensing model (R. Ivison, private communication).  A fit to the $uv$ 
visibilities yields FWHM diameters of $3.2\arcsec$ (east-west) and $\leq 
0.8\arcsec$ (north-south); correcting the former for magnification by a 
factor of ${\cal M} = 2.45$ implies a source-plane linear size of $\sim 
10\,h_{0.7}^{-1}\,{\rm kpc}$.  An additional tail of emission extending 
$\geq 4\arcsec$ WSW from the central peak may be associated 
with extended ${\rm Ly\,\alpha}$ emission detected by \citet{vern01}, and with 
another fainter $R$-band emission peak (Figure \ref{f-cont}).  The $\sim 
10\,h_{0.7}^{-1}\,{\rm kpc}$ size of the central dust source in J02399 is 
about three times larger than the sizes of local ULIRGs \citep{down98,brya99}.
This fact may not be surprising, however, given that J02399 is about ten 
times more luminous than the typical local ULIRG.  Since the star-forming 
interstellar medium in ULIRGs is thought to consist of a high volume 
filling factor distribution of dense gas and dust \citep{down98,saka99}, 
a ten times larger star formation rate requires a ten times larger volume 
of star-forming regions.  

Table \ref{t-posn} summarizes the absolute positions and astrometric errors of 
the emission peak of J02399 in the different wavebands.  The coincidence of 
the far-IR centroid with the UV and X-ray continuum peaks within the combined 
astrometric uncertainties of $\sim 1\arcsec$ suggests that L1 is indeed the 
primary origin of J02399's dust emission.  The (bluer) UV continuum and ${\rm 
Ly\,\alpha}$ emission peak L2, in contrast, has no associated dust emission; 
this supports the idea that L2 is primarily scattered AGN light from L1 
\citep{vern01}, rather than a second galaxy in the early stages of a merger 
with L1 \citep{ivis98}.   In fact, there may be an anticorrelation between the 
fainter rest UV emission and the $335\,{\rm \mu m}$ continuum emission, 
suggesting that extinction strongly affects the short-wavelength morphology of 
the source.  UV emission may only escape from ``extinction holes'' in the 
extremely dusty environment of J02399, a scenario also suggested by 
\citet{ivis01} in their study of the dusty high-redshift SCUBA source 
SMM\,J14011+0252 (hereafter J14011).

\subsection{CO line emission}

Figure \ref{f-mom0} shows the velocity-integrated CO(3--2) line emission 
($\lambda_{\rm obs} = 3.30\,{\rm mm}$) at a resolution of $5.2\arcsec \times 
2.4\arcsec$ in greyscale, with the continuum contours of Figure \ref{f-cont} 
overlaid.  Taking into account the different spatial resolutions, the 
distributions and centroids of the velocity-integrated CO(3--2) and 
$335\,{\rm \mu m}$ continuum emission are coincident, and the peak CO emission 
(like the peak continuum) originates in the eastern (blue-shifted) emission 
peak.  The UV/X-ray source L1 is located $0.7\arcsec \pm 0.6\arcsec$ from the 
rotation center.  Figure \ref{f-line} shows the CO(3--2) line profile, smoothed
to $60\,{\rm km\,s^{-1}}$ and integrated over the central $\sim 5\arcsec$.
Figure \ref{f-chan} displays channel maps of the CO(3--2) emission (spaced by 
$200\,{\rm km\,s^{-1}}$) at a resolution of $5.2\arcsec \times 2.4\arcsec$.  
Our new data agree with and significantly improve on the earlier $7\arcsec 
\times 5\arcsec$ interferometry of \citet{fray98}.  The molecular line 
emission of J02399 is very broad and has a symmetric ``double-horn'' profile 
which is characteristic of rotation.  Line emission extends over more than 
$1100\,{\rm km\,s^{-1}}$ (FWHM), and the global profile shows emission peaks 
at $-520\,{\rm km\,s^{-1}}$ and $+230\,{\rm km\,s^{-1}}$ (relative to the 
assumed $z_{\rm CO} = 2.808$).  An improved CO systemic redshift for the 
source would thus be $2.8076 \pm 0.0002$.  Channel maps show that blue and 
red-shifted emission maxima originate $1.5\arcsec - 2\arcsec$ on either side 
of the phase center, again consistent with rotation.  The largest velocity 
gradient is along a position angle $\sim 115^\circ$, i.e., close to the 
direction of the elongation of the continuum emission.  In addition to this 
regular behavior, there also appears to be highly blueshifted emission (in the 
$-600\,{\rm km\,s^{-1}}$ channel) northwest of the center.  We interpret the 
line profiles and channel maps as showing rapid rotation about a position very 
near the phase center.  Taken together, the line and continuum data indicate 
that the most likely morphology of the source is a disk of molecular gas and 
dust rotating about a central AGN located at the position of the hard X-ray 
peak.  The absolute astrometry renders unlikely but does not exclude an 
alternative configuration of two galaxies orbiting each other, in which the 
AGN coincides with either the red or the blue-shifted CO(3--2) emission peak.

\section{Discussion}\label{s-disc}

\subsection{J02399 is a very massive galaxy}\label{ss-mass}

In what follows, we assume that the molecular gas in J02399 is indeed 
distributed in a rotating disk.  We have constructed ring/disk models with 
flat rotation curves to fit the continuum and line profile data, including the 
effects of beam smearing and nonzero velocity dispersions.  The data are well 
fit (Figures \ref{f-line} and \ref{f-disk}) by a disk model with a (radially 
constant) rotation velocity of $420 \pm 20\,{\rm km\,s^{-1}}$ for an assumed 
inclination of $i = 90^\circ$.  The broad wings of the integrated line profile 
indicate that large local random motions (turbulence) are present in the disk. 
The fit requires a ratio of local velocity dispersion to rotation velocity of 
$\sim 0.23$, which is $\sim 2-3$ times larger than in the Milky Way disk but 
comparable to that seen in local ULIRG mergers \citep{down98}.  A ring-like 
distribution with a maximum gas density at $r \sim 1\arcsec$ on the sky ($R = 
3.2\,h_{0.7}^{-1}\,{\rm kpc}$ in the source plane) and $\Delta R({\rm FWHM})/R 
\sim 1-1.5$ provides a better match to the flux ratio of $\sim 2$ between the 
``horns'' and the central dip in the global spectrum than a filled, 
centrally-peaked disk.  \citet{down98} have found that similar rings on 
somewhat smaller scales are characteristic of the molecular gas distributions 
in local ULIRGs.  The outer disk radius required by our model is $r \sim 
2\arcsec - 2.5\arcsec$ ($R \sim 6-8\,h_{0.7}^{-1}\,{\rm kpc}$), in good 
agreement with our direct fit to the $335\,{\rm \mu m}$ continuum source size. 
The enclosed dynamical mass inferred from this disk model is 
\begin{equation}
{\frac {M_{\rm dyn}}{M_\odot}} = 1.3 \times 10^{11}\,\gamma\,\Big({\frac {r}
{1\arcsec}}\Big)\,\Big({\frac {D_{\rm A}}{1.6\,h_{0.7}^{-1}\,{\rm 
Gpc}}}\Big)\,\Big({\frac {\cal M}{2.45}}\Big)^{-1}\,\Big({\frac {v_{\rm rot}}
{420\,{\rm km\,s^{-1}}}}\Big)^2\,\Big({\frac {{\rm sin}\,i}{1.0}}\Big)^{-2}
\end{equation}
Here $\gamma \leq 1$ is a dimensionless scale factor which depends on the 
disk/spheroid structure, and which for a thin disk in a massive spheroid is 
close to unity \citep{lequ83}.  Within the $r \sim 2.5\arcsec$ outer radius of 
the molecular emission, the dynamical mass is $\sim 3.2 \times 
10^{11}\,h_{0.7}^{-1}\,M_\odot$.  If a merger model is used instead of 
a rotating disk, the enclosed mass is approximately a factor of two larger.  

The integrated CO(3--2) flux from J02399 is $3.1 \pm 0.4\,{\rm 
Jy\,km\,s^{-1}}$ ($\pm 0.3\,{\rm Jy\,km\,s^{-1}}$ for a 10\% uncertainty in 
the flux scale), in excellent agreement with the value of $3.0\,{\rm 
Jy\,km\,s^{-1}}$ found by \citet{fray98}.  With this integrated flux and the 
assumption of moderately dense ($n_{\rm H_2} \gg 10^3\,{\rm cm^{-3}}$), 
thermalized gas at temperature $\gtrsim 30\,{\rm K}$ (see e.g., Combes, Maoli, 
\& Omont 1999), the molecular (plus helium) gas mass can be estimated 
following \citet{solo97}:
\begin{equation}
{\frac {M_{\rm gas}}{M_\odot}} = 1.60 \times 10^{10}\,\alpha\,\Big({\frac 
{F_{\rm CO}}{\rm Jy\,km\,s^{-1}}}\Big)\,\Big({\frac {\cal M}
{2.45}}\Big)^{-1}\,\Big({\frac {\nu_{\rm obs}}{90.81\,{\rm 
GHz}}}\Big)^{-2}\,\Big({\frac {D_{\rm L}}{23.5\,h_{0.7}^{-1}\,{\rm 
Gpc}}}\Big)^2\,\Big({\frac {1+z}{3.808}}\Big)^{-3}
\end{equation}
Here $\alpha$ is the conversion factor from CO luminosity to gas mass, which 
we assume to be $\sim 0.8-1.6\,M_\odot\,({\rm K\,km\,s^{-1}\,pc^2})^{-1}$, as 
estimated for local ULIRGs \citep{solo97,down98}.  The inferred 
molecular gas mass in J02399 is thus $6.0 \pm 2.4 \times 10^{10}\,
h_{0.7}^{-2}\,M_\odot$, implying a gas mass fraction of roughly 10--30\% 
within $r = 2.5\arcsec$ which is similar to those observed in local ULIRGs 
\citep{down98}. 

J02399 is clearly a very massive system, and is in particular a very massive 
{\it baryonic} system.  Observational studies of local ellipticals (Keeton 
2001; Boriello, Salucci, \& Danese 2002; cf. Loewenstein \& Mushotzky 2002) 
suggest that typically $\sim 30\%$ of the total mass within the optical 
effective radius will be dark; the interior regions of massive local disk 
galaxies have similarly low dark matter fractions \citep{comb02}.  By analogy 
with these local systems, J02399 should have a baryonic mass of $\geq 2 \times 
10^{11}\,{\rm sin}^{-2}\,i\,h_{0.7}^{-1}\,M_\odot$ within $8\,h_{0.7}^{-1}\,
{\rm kpc}$; $M_{\rm gas} \sim 6 \times 10^{10}\,h_{0.7}^{-2}\,M_\odot$ 
then implies that a conservative 
lower limit to the stellar mass within the same radius is $\sim 1.4 \times 
10^{11}\,M_\odot$ (for $i = 90^\circ$).  The \citet{kenn83} stellar initial 
mass function (IMF) models of \citet{cole01}\footnote{The \citet{kenn83} IMF 
is comparable in this context to a \citet{salp55} IMF from $m_{\rm lower} = 
1\,M_\odot$ to $m_{\rm upper} = 100\,M_\odot$.  For a $0.1-100\,M_\odot$ 
Salpeter IMF, \citet{cole01} find $m^*(z = 0) = 1.4 \times 
10^{11}\,h_{0.7}^{-2}\,M_\odot$.} imply that $m^* \sim 7 \times 
10^{10}\,h_{0.7}^{-2}\,M_\odot$ for the galaxy stellar mass function in the 
local Universe.  Based on the stellar content of its inner 
$8\,h_{0.7}^{-1}\,{\rm kpc}$, then, 
J02399 is already at least a $2m^*$ system, on its way to becoming a $\geq 
3m^*$ system as it turns the remainder of its molecular gas into stars.

Our new high-resolution observations of J02399 provide a fairly reliable 
estimate of the dynamical mass of a massive high-$z$ galaxy.  The more easily 
available observations of the ionized gas kinematics are more prone to 
systematic bias and influences other than gravity \citep{pett01,heck00}.  
The presence of such a massive object at a redshift as high as $\sim 3$ is 
qualitatively surprising, given that CDM models make the general prediction 
that galaxy masses build up gradually as a function of time and 
redshift, such that the comoving density of massive objects at $z \sim 2$ to 3 
is only a small fraction of their local density (e.g., Kauffmann \& Charlot 
1998; Baugh et al. 2002).  However, given the small number of objects
whose kinematics have so far been studied with millimeter interferometry, the 
question arises whether the detection of a few such massive objects is 
significant, or whether they are mere items of curiosity.  

Besides J02399, the SCLS has yielded another source whose high redshift has 
been confirmed by spatially resolved CO spectroscopy.  J14011 at $z_{\rm CO} = 
2.565$ has an observed $850\,{\rm \mu m}$ flux density (12\,mJy), a lensing 
magnification (2.5: Ivison et al. 2001), a bolometric luminosity ($1.3 \times 
10^{13}\,h_{0.7}^{-2}\,L_\odot$), and a molecular gas mass ($3.4 \times 
10^{10}\,h_{0.7}^{-2}\,M_\odot$) which are similar to those of J02399 
\citep{fray99,ivis01}.  The CO line width of J14011 (FWHM $\Delta v = 
190\,{\rm km\,s^{-1}}$: Frayer et al. 1999; Downes \& Solomon 2002) is much 
narrower than that of J02399 (FWHM $\Delta v = 1100\,{\rm km\,s^{-1}}$), 
however.  Although no clear rotation pattern has as yet been detected and no 
inclination correction can be made on the basis of the available CO line and 
millimeter continuum maps \citep{fray99,down02}, the inclination must be less 
than $20^\circ$ for the dynamical mass to be $\geq 7 \times 
10^{10}\,h_{0.7}^{-1}\,M_\odot$ (i.e., at least twice the gas mass).  If the 
complex UV source structure is taken as evidence of a merger morphology, the 
dynamical mass would be $\geq 1.4 \times 10^{11}\,h_{0.7}^{-1}\,M_\odot$.  
\citet{down02} propose a disk model with a rotation velocity of $250\,{\rm 
km\,s^{-1}}$, an inclination of $50^\circ$, and an outer radius of 
$1.1\arcsec$; for these parameters and magnification by a factor ${\cal M} = 
2.5$, J14011's dynamical mass is $6 \times 10^{10}\,h_{0.7}^{-1}\,M_\odot$.  
Their further suggestion that J14011 undergoes an additional factor of ten 
magnification by a previously unrecognized foreground galaxy (and that $M_{\rm 
dyn}$ is therefore a factor of ten lower) rests on the assumption 
that the filling factor of CO emission within the $1.1^{\prime\prime}$-radius 
structure is unity.  Of the seven background SCLS sources with intrinsic 
$S_{850} \geq 4\,{\rm mJy}$ (before lensing magnification), then, the two with 
known redshifts $\geq 2$ both have likely baryonic masses in the $\gtrsim 
10^{11}\,M_\odot$ bin.  These source parameters are not atypical at high 
redshift: in a growing number of high-$z$ QSOs and radio galaxies, millimeter 
CO interferometry reveals comparably large molecular gas masses (in several 
cases already $\geq 10^{11}\,h_{0.7}^{-2}\,M_\odot$) and a variety of 
linewidths (Guilloteau et al. 1999; Papadopoulos et al. 2000; Cox et al. 2002, 
and references therein).

\subsection{Cosmic volume densities of massive galaxies as a function of 
redshift}\label{ss-volu}

For our cosmology, the observed area covered by the SCLS ($3.6 
\times 10^{-6}\,{\rm sr}$) and the $1 \leq z \leq 5$ range sampled by the 
$850\,{\rm \mu m}$ observations correspond to a comoving volume of 
$5.3 \times 10^5\,h_{0.7}^{-3}\,{\rm Mpc}^3$.  
The effective comoving volume density corresponding to one source like J02399 
is thus $\Phi(M \geq 10^{11}\,M_\odot) = 10^{-5.7}\,h_{0.7}^3\,{\rm 
Mpc}^{-3}$.  Including J14011, this becomes $10^{-5.4}\,h_{0.7}^3\,{\rm 
Mpc}^{-3}$; including the four remaining SCLS sources with intrinsic $S_{850} 
\geq 4\,{\rm mJy}$ and unknown redshifts (on the assumption that they are also 
at $z \geq 1$ and comparably massive) boosts $\Phi$ to 
$10^{-4.95}\,h_{0.7}^3\,{\rm Mpc}^{-3}$.  Although the latter step is the most 
uncertain, evidence from the weakness of radio and optical/UV counterparts for 
submillimeter sources suggests that these additional four sources are more 
likely to lie at high than at low redshifts \citep{dann02,ivis02}.  Regardless 
of whether we assign one, two, or six SCLS sources to the $\geq 
10^{11}\,M_\odot$ baryonic mass bin, however, we must adjust $\Phi$ to reflect 
the fact that not all massive objects will be detectable by SCUBA as 
$\sim 10\,{\rm mJy}$ sources at all times.  For a Miller-Scalo (1979) IMF, or 
a $1-100\,M_\odot$ Salpeter (1955) IMF, the bolometric (infrared) luminosity 
and star formation rate are related by \citep{kenn98}
\begin{equation}
{\frac {\rm SFR}{M_\odot\,{\rm yr^{-1}}}} \sim {\frac {L_{\rm IR}}{1.2 \times 
10^{10}\,L_\odot}}
\end{equation}
such that the star formation rate in J02399 (for an assumed AGN contribution 
to $L_{\rm IR}$ of 50\%) is about $500\,M_\odot\,{\rm yr^{-1}}$.  Turning a 
total baryonic mass of $\sim 2 \times 10^{11}\,M_\odot$ into stars at this 
rate will require a total of $\sim 4 \times 10^8\,{\rm yr}$. 
Since the time elapsed over the range $1 \leq z \leq 5$ is $\sim 4.6\,{\rm 
Gyr}$, correction for the finite starburst lifetime will increase the total 
$\Phi (M \geq 10^{11}\,M_\odot)$ by at least a factor of $\sim 11.5$ over our 
estimates above.  The correction factor can be higher still if any of the 
past star formation in J02399 occurred in a quiescent mode, if the current 
burst will be terminated prematurely by strong negative feedback to the 
interstellar medium (e.g., Genzel \& Cesarsky 2000), or if the dark matter 
fraction within $R = 8\,h_{0.7}^{-1}\,{\rm kpc}$ is greater than 30\%.

We now consider the evolution of $\Phi(M \geq 10^{11}\,M_\odot)$ with redshift.
In the upper panel of Figure \ref{f-sams}, we have plotted at $z \sim 2-3$ the 
comoving volume densities derived for two and six SCLS sources having baryonic 
masses $\geq 10^{11}\,M_\odot$; upward arrows indicate a factor of 11.5 
starburst lifetime correction.  At $z \sim 1$, we have plotted the volume 
densities of $K$-selected systems with stellar masses $\geq 10^{11}\,
h_{0.7}^{-2}\,M_\odot$ as inferred from a population synthesis analysis 
of Munich Near-Infrared Cluster Survey (MUNICS) sources by \citet{dror02}.  We 
also show the volume densities of actively star-forming galaxies ($S_{15} 
\geq 300\,{\rm \mu Jy}$) detected in deep ISOCAM surveys \citep{elba99,genz00}.
\citet{rigo02} have obtained ${\rm H\alpha}$ rotation curves for three of 
these systems and find dynamical masses $\geq 10^{11}\,h_{0.7}^{-1}\,M_\odot$.
Assuming that all ISOCAM galaxies with $15\,{\rm \mu m}$ flux densities 
$\geq 300\,{\rm \mu m}$ in the redshift range $0.2 \leq z \leq 1.3$ (sampled 
by the spectroscopy of Rigopoulou et al. 2002) have such masses, we derive a 
volume density of $2.4 \times 10^{-4}\,h_{0.7}^3\,{\rm Mpc}^{-3}$.  
At $z = 0$, we show the comoving volume density of systems with stellar masses 
$\geq 10^{11}\,h_{0.7}^{-2}\,M_\odot$ as obtained from modelling optical and 
near-IR 2dF and 2MASS photometry \citep{cole01}.  

For comparison with the observational data points, we also plot curves showing 
the predictions of two different semi-analytic models for galaxy evolution 
within a $\Lambda$CDM structure formation cosmogony.  For a representative 
``Munich'' model, we count the number of galaxies with total stellar+gas mass 
$\geq 10^{11}\,M_\odot$ in the catalogues of \citet{kauf99} and divide by the 
simulation volume of $(200\,h_{0.7}^{-1}\,{\rm Mpc})^3$; the comoving 
densities are marginally higher than those we would derive from counting only 
systems with stellar mass $\geq 10^{11}\,M_\odot$.  For a representative 
``Durham'' model, we use a curve for stellar mass $\geq 10^{11}\,M_\odot$ 
(C. Baugh, private communication) derived from a recent variation on the 
reference model of \citet{cole00}.  Although both Munich and Durham models 
adopt similar input parameters for their derivations of dark matter halo 
distributions ($\Omega_{\rm M} = 0.3$, $\Omega_\Lambda = 0.7$, $h_{0.7} = 1$, 
$\sigma_8 \simeq 0.9$, and $\Gamma \simeq 0.2$), they treat baryons 
differently.  \citet{kauf99} adopt a higher cosmic baryon density 
($\Omega_{\rm b} = 0.045$ vs. 0.02), and the two groups use different recipes 
for star formation, feedback, and other physical processes.  As a result, the 
models' predictions for the rate of assembly of systems with baryonic masses 
$\geq M_{\rm min} = 10^{11}\,M_\odot$ differ substantially at high redshifts, 
as previously noted by Benson, Ellis, \& Menanteau (2002)\footnote{See in 
particular Figure 1 of astro-ph/0110387.}.  Consideration of observational 
data (note that $M_{\rm min}$ scales differently with $h_{0.7}$ for the 
different sets of sources) begins to make it possible to discriminate between 
the models.  We find that the volume densities of massive SCUBA galaxies are 
clearly above the predictions of the Durham model, and are significantly above 
the predictions of the Munich model if a reasonable correction for a finite 
starburst lifetime is applied.  At $z \sim 1$, the data are still above the 
Durham model, but in good agreement with the Munich model.  Both the Durham 
and the Munich models account well for the local volume densities of massive 
galaxies.

\subsection {Comparison to QSOs}

J02399 is a type 2 QSO.  It is thus of interest to compare our estimated 
volume densities for the brightest submillimeter galaxies to those of 
high-$z$ QSOs.  Converting the UV/optical and X-ray luminosities of type 1 
QSOs to bolometric luminosities with the \citet{elvi94} correction factors and 
assuming that the central black hole contains about 0.15\% of the stellar 
spheroid mass \citep{ho99,korm01}, the QSO host mass is related to the 
optical/X-ray 
luminosity by 
\begin{equation}
{\frac {M_{\rm host}}{M_\odot}} = 10^{10}\,\Big({\frac {f_\nu\,L_\nu}{4.9 
\times 10^{11}\,L_\odot}}\Big)\,\eta_{\rm Edd}^{-1}
\end{equation}
Here $f_\nu$ is the bolometric correction for the band luminosity $L_\nu$ 
($f_{\rm X} \sim 30$ and $f_{\rm opt} \sim 14$: Elvis et al. 1994), and 
$\eta_{\rm Edd}$ is the radiative efficiency relative to the Eddington 
rate.\footnote{Note that this relationship predicts a stellar spheroid mass 
for J02399 of $\geq 2.4 \times 10^{11}\,h_{0.7}^{-2}\,M_\odot$ if 50\% of its 
$L_{\rm IR}$ comes from a buried type 2 QSO, $\eta_{\rm Edd} \sim 0.5$, 
and $f_{\rm IR} \geq 1$.}  For $\eta_{\rm Edd} \sim 0.5$, host masses 
$\geq 10^{11}\,h_{0.7}^{-2}\,M_\odot$ correspond to X-ray luminosities $\geq 3 
\times 10^{44}\,h_{0.7}^{-2}\,{\rm erg\,s^{-1}}$ and absolute $B$, $V$, and 
$R$ magnitudes $\leq -24.7$.    From the X-ray work of Miyaji, Hasinger, \& 
Schmidt (2001) and the optical work of \citet{hawk96}, we then estimate the 
density of type 1 AGN with $M_{\rm host} \geq 10^{11}\,h_{0.7}^{-2}\,M_\odot$ 
in the range $1.5 \leq z \leq 3$ to be $\sim 10^{-6 \pm 0.5}\,h_{0.7}^3\,
{\rm Mpc}^{-3}$.  The QSO lifetime $t_{\rm QSO}$ may be a factor of three 
shorter than $t_{\rm starburst}$ \citep{nuss93}, and the ratio of type 2 to 
type 1 QSOs is $\rho \geq 4$ (Gilli, Salvati, \& Hasinger 2001).  Correcting 
for $t_{\rm starburst}/t_{\rm QSO} \times \rho \sim 12$, the comoving density 
of massive QSO hosts is $\sim 10^{-5 \pm 0.5}\,h_{0.7}^3\,{\rm Mpc}^{-3}$, in 
remarkably-- and probably fortuitously-- good agreement with the values 
estimated for the bright SCUBA sources listed above.  This agreement is 
consistent with the general assumption that the evolution of the most massive 
black holes and their host galaxies is connected through common formation at 
high redshift.

\subsection{Is the discrepancy between observed and predicted volume densities 
mass-dependent?}

CDM models do very well in accounting for the large-scale structure, luminosity
functions, and color distributions of galaxies at $z = 0$ (e.g., Kauffmann et 
al. 1999; Cole et al. 2000; papers in Guiderdoni et al. 2001).  They predict 
the continuous buildup of galaxy masses from small to large through the basic 
process of hierarchical merging.  If our observations indicate that these 
models underpredict the high-$z$ volume densities of massive galaxies, the 
next question is whether this shortfall originates in the predicted evolution 
of the underlying $\Lambda$CDM 
halo distributions, in some general feature of the semi-analytic models which 
build on them, or in some specific aspect of how the semi-analytic recipes 
treat baryonic processes in the most massive systems.  Consideration of the 
halo catalogues of \citet{kauf99} offers no obvious evidence that the 
parameters of the $\Lambda$CDM simulations per se are at fault. If we multiply 
each halo mass by $\Omega_{\rm b}\,(\Omega_{\rm M} - \Omega_{\rm b})^{-1}$ to 
estimate the associated baryonic mass, then even for the lower $\Omega_{\rm b} 
= 0.02$ preferred by the Durham group, the comoving volume densities of haloes 
with {\it available} baryonic mass $\geq 10^{11}\,M_\odot$ comfortably exceed 
the volume densities of galaxies with {\it observed} baryonic mass $\geq 
10^{11}\,M_\odot$ at every redshift.  This statement remains true even if the 
point plotted for six SCLS sources in the upper panel of Figure \ref{f-sams} 
gains a full factor of $\sim 11.5$ correction for finite starburst lifetimes.  

To evaluate the performance of the semi-analytic models in a lower mass 
regime, we compare in the lower panel of Figure \ref{f-sams} the model 
predictions with the comoving volume densities of $z \sim 3$ Lyman break 
galaxies (LBGs).  These 
systems have average stellar masses near but somewhat below the knee of the 
Schechter function ($M \sim 10^{10}\,h_{0.7}^{-2}\,M_\odot \sim 0.14\,m^*$), 
according to the recent stellar synthesis analyses of Papovich, Dickinson, \& 
Ferguson (2001) and \citet{shap01}.  Again, we compare these measurements to 
the Durham and Munich model predictions.  In this lower mass bin, the 
semi-analytic models still fall below (but now closer to) the observed volume 
densities of LBGs.  In addition, we have plotted the model predictions of 
Somerville, Primack, \& Faber (2001), who include the effects of extinction on 
the predicted volume densities for $R < 25.5$ LBGs and consider two extreme 
star formation histories-- one with a constant star formation timescale, and 
one including star formation bursts triggered by galaxy collisions and 
mergers.  Clearly the low star formation rates predicted by the constant model 
fail by a large factor in accounting for the data, while the burst models fit 
the data very well.  The Munich and Durham models (which include some 
acceleration of the star formation rate at high redshift) lie between 
these two extremes.  It appears that the recipes for star formation and 
feedback used in the semi-analytic models do reasonably well in accounting for 
sub-$m^*$ galaxies, once the influence of merger-induced bursts is included.

Our discussion thus suggests that the discrepancy between models and data may 
increase with galaxy mass.  Systems with masses a few times $m^*$ 
have comoving number densities at $z \sim 1$ which are comparable to those 
in the local Universe ($z = 0$).  At $z = 2-3$, the same kinds of objects 
still have comoving volume densities which are $\geq 10\%$ of their $z = 0$ 
values.  These findings appear to contradict current hierarchical CDM 
models, which predict a large decrease of the abundance of massive galaxies 
at $z \geq 1$ due to an extended phase of merger activity (e.g., Kauffmann \& 
Charlot 1998).  The data presented here indicate that this picture may 
have to be revised with respect to the rate at which massive systems (and 
bulges) are assembled at fairly high redshift, possibly through adjustment 
of $\Omega_{\rm b}$ and its coupling to very efficient star formation in major 
mergers.  More work is needed before a firmer statement can be made about this 
crucial, ``baryonic mass assembly'' test of the current standard cosmogony.  
Both observational and theoretical estimates are still very uncertain, with 
the former depending on small samples and large lifetime corrections, and the 
latter on ad hoc input recipes for star formation and feedback.  

\acknowledgments

We thank IRAM staff members, particularly Roberto Neri, for their help in 
coordinating and conducting the observations for this project, and Rob Ivison
for sharing information about the SCLS lensing model.  We are especially 
grateful to Len Cowie for providing us with his high-resolution Keck images 
and astrometry of J02399 prior to publication.  We also thank Carlton Baugh 
for providing the most current Durham model predictions, Mark Bautz for 
clarifying the Chandra astrometry, and Chung-Pei Ma, Chris McKee, and Ian 
Smail for very useful comments.  This research has made use of the 
NASA/IPAC Extragalactic Database (NED), which is operated by the Jet 
Propulsion Laboratory, California Institute of Technology, under contract with 
NASA.

\clearpage

\begin{deluxetable}{ccllcl}
\tablecaption{Positions and uncertainties for J02399 in different wavebands.
\label{t-posn}}
\tablewidth{0pt}
\tablehead{
\colhead{Band} & 
\colhead{} &
\colhead{R.A.} & 
\colhead{Dec.} & 
\colhead{$1\sigma$ astrometric} & 
\colhead{} \\
\colhead{(rest-frame)} & 
\colhead{Feature} &
\colhead{(J2000)} & 
\colhead{(J2000)} & 
\colhead{uncertainty} & 
\colhead{Reference} 
}
\startdata
$335\,{\rm \mu m}$ & centroid & 02:39:51.87 & --01:35:58.8 & $\pm 0.6\arcsec$ & 1 \\
$1685\,{\rm \AA}$ & peak (L1) & 02:39:51.845 & --01:35:58.2 & $\pm 0.3\arcsec$ & $2^a$ \\
0.4--1.8\,keV & peak & 02:39:51.81 & --01:35:58.1 & $\pm 0.6\arcsec$ & $3^b$ \\
\enddata
\tablenotetext{a}{Based on an astrometric solution of USNO-A2 stars relative 
to VLA sources in the Abell 370 field.}
\tablenotetext{b}{Includes a $(\Delta\,{\rm R.A.}, \Delta\,{\rm Dec}) = 
(-1.6\arcsec,+0.5\arcsec)$ overall ``boresight'' correction between Chandra 
and APM optical reference frames.}
\tablerefs{(1) this paper (2) L. Cowie, private communication (3) 
\citet{baut00}; M. Bautz, private communication}
\end{deluxetable}

\clearpage

\figcaption{
Contours of 1.27\,mm (rest-frame $335\,{\rm \mu m}$) continuum emission from 
J02399 mapped with the PdBI, superposed on a greyscale image of the $R$-band 
(rest-frame UV) continuum obtained at Keck (L. Cowie, private communication).  
Contours are multiples of $0.4\,{\rm mJy\,beam^{-1}}$, for a synthesized beam 
of $1.8\arcsec \times 1.4\arcsec$ at position angle $42^\circ$ east of north 
(shown at lower left).  Arrows indicate rest-UV peaks L1 and 
L2; cross denotes the position and $1\sigma$ positional uncertainty (including 
both X-ray and millimeter astrometric uncertainties) of the hard X-ray source  
(Bautz et al. 2000; M. Bautz, private communication).  The absolute 
uncertainty of the PdBI astrometry is $\pm 0.6\arcsec$; the $R$-band 
astrometry is based on a comparison of USNO-A2 stars to VLA sources in the 
Abell 370 field and has a relative uncertainty of $\pm 0.3\arcsec$; the X-ray 
astrometry has an uncertainty of $\pm 0.6\arcsec$.
\label{f-cont}}

\figcaption{Overlay of continuum (contours, as in Figure \ref{f-cont})
and CO(3--2) velocity-integrated line emission (greyscale).  The CO(3--2) map 
has a synthesized beam of $5.2\arcsec \times 2.4\arcsec$ at position angle 
$33^\circ$ east of north.
\label{f-mom0}}

\figcaption{CO(3--2) spectrum, smoothed to $60\,{\rm km\,s^{-1}}$ resolution 
and integrated over the central $\sim 5\arcsec$ of the source.  Per-channel 
rms is indicated at lower left.  Velocities are with respect to the assumed 
redshift $z_{\rm CO} = 2.808$ \citep{fray98}.  The dashed curve is a model 
spectrum, computed for a rotating disk with constant rotation velocity $v_{\rm 
rot} = 420\,{\rm km\,s^{-1}}$, local FWHM random motions of $220\,{\rm 
km\,s^{-1}}$, inclination $i = 90^\circ$, and a Gaussian radial intensity 
distribution which peaks at $1.0\arcsec$ and has radial FWHM $1.5\arcsec$ (see 
\S \ref{ss-mass}).  The FWHM vertical thickness of the disk/ring is assumed to 
be $0.5\arcsec$.  The model takes into account spatial and velocity smearing.
\label{f-line}}

\figcaption{
CO(3--2) channel maps in steps of $200\,{\rm km\,s^{-1}}$ with $0.35\,{\rm 
mJy\,beam^{-1}}$ contour spacing.  Velocities are with respect to the assumed 
redshift $z_{\rm CO} = 2.808$ \citep{fray98}.  The synthesized beam 
(shown at lower left) is $5.2\arcsec \times 2.4\arcsec$ at position angle 
$33^\circ$ east of north.
\label{f-chan}}

\figcaption{
Position-velocity diagram of observed CO(3--2) emission, at $60\,{\rm 
km\,s^{-1}}$ resolution and position angle $115^\circ$ (white contours are 
multiples of $0.55\,{\rm mJy\,beam^{-1}}$), overlaid on the predictions of 
the rotating disk model (greyscale) described in \S \ref{ss-mass} and in the 
caption for Figure \ref{f-line}.
\label{f-disk}}

\figcaption{
Cosmic densities of galaxies of (baryonic/stellar) mass $\geq M_{\rm 
min}$ for the flat $\Omega_\Lambda = 0.7$, $H_0 = 70\,h_{0.7}\,{\rm 
km\,s^{-1}\,Mpc^{-1}}$ cosmology used throughout. {\bf Upper panel:} 
$M_{\rm min} = 1 \times 10^{11}\,M_\odot$.  The filled dark circles 
(and $1\sigma$ error bars) are estimated from the SCLS galaxies J02399 
and J14011 (lower), and from 
the six SCLS galaxies with intrinsic $S_{850} \geq 4\,{\rm mJy}$ which may 
lie at $z \geq 1$ (upper).  Mass estimates are from the CO kinematics in 
J02399 and J14011 (this paper; Ivison et al. 2001; Downes \& Solomon 2002).  
The dashed upward arrows indicate correction factors for finite starburst 
lifetimes, as discussed in \S \ref{ss-volu}.  The light filled square 
denotes the density of $S_{15} \geq 300\,{\rm \mu Jy}$ ISOCAM galaxies in the 
range $z = 0.2-1.3$, for three of which \citet{rigo02} have determined large
dynamical masses from ${\rm H\alpha}$ rotation curves.  The open rectangles 
with crosses are source densities for massive early-type galaxies from the 
MUNICS $K$-band work of \citet{dror02}, where stellar masses are estimated 
from stellar synthesis models.  The filled triangle and bar denote the $z = 0$ 
results obtained from stellar synthesis modelling of 2dF/2MASS data by 
\citet{cole00}.  For comparison, we show two theoretical predictions from 
semi-analytic models in CDM cosmogonies: the continuous curve is from a 
``Durham'' model (C. Baugh, private communication; see also Cole et al. 2001; 
Baugh et al. 2002), while the dashed curve is from a ``Munich'' model 
\citep{kauf99}.  In all cases, the stellar masses refer to a Miller-Scalo 
(1979) IMF, or a Salpeter (1955) IMF between 1 and $100\,M_\odot$.  {\bf Lower 
panel:} $M_{\rm min} = 10^{10}\,M_\odot$.  Estimates of the density of $z \sim 
3$ Lyman break galaxies with $M \geq M_{\rm min}$ (as determined from stellar 
synthesis modelling) are denoted by a filled rectangle \citep{shap01} and 
downward-pointing light triangle \citep{papo01}.  All other notation is as in 
the upper panel.  Thin and thick short-dashed curves are models from the 
work of \citet{some01}, which take into account dust extinction and represent 
constant and merger-induced (``collisional'') star formation histories, 
respectively.
\label{f-sams}}


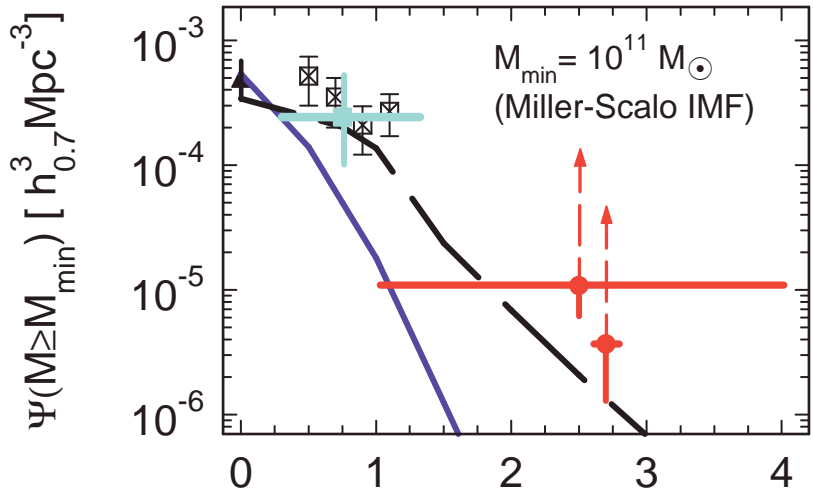
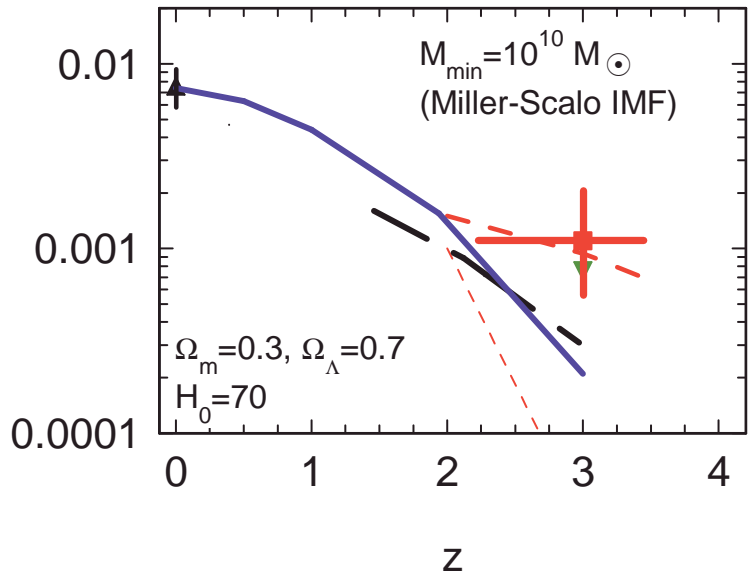
\end{document}